\documentclass[12pt,a4paper]{iopart}
\usepackage[final]{graphicx}
\usepackage{siunitx}
\usepackage{float}
\usepackage{braket}
\usepackage[caption = false]{subfig}

\usepackage{color}
\usepackage{soul}
\usepackage{ulem}

\begin{document}

\title{Suppressing phonon decoherence of high performance single-photon sources in nanophotonic waveguides   }

\author{C. L. Dree{\ss}en$^1$, C. Ouellet-Plamondon$^1$, P. Tighineanu$^{1,2}$, X. Zhou$^1$, L. Midolo$^1$, A. S. S{\o}rensen$^1$, and P. Lodahl$^1$}

\address{$^1$ Hy-Q Center for Hybrid Quantum Networks, Niels Bohr Institute, University of Copenhagen, Blegdamsvej 17, 2100 Copenhagen, Denmark\\
$^2$ Max Planck Institute for the Science of Light, Staudtstra{\ss}e 2, 91058 Erlangen, Germany}
\ead{chris.dreessen@nbi.ku.dk, lodahl@nbi.ku.dk}
\vspace{10pt}
\begin{indented}
\item[]June 2018
\end{indented}

\begin{abstract}
The fundamental process limiting the coherence of quantum-dot based single-photon sources is the interaction with phonons. We study the effect of phonon decoherence on the indistinguishability of single photons emitted from a quantum dot embedded in a suspended nanobeam waveguide. At low temperatures, the indistinguishability is limited by the coupling between the quantum dot and the fundamental vibrational modes of the waveguide and is sensitive to the quantum-dot position within the nanobeam cross-section. We show that this decoherence channel can be efficiently suppressed by clamping the waveguide with a low refractive index cladding material deposited on the waveguide. With only a few microns of cladding material, the coherence of the emitted single photons is drastically improved. We show that the degree of indistinguishability can reach near unity and become independent of the quantum-dot position. We finally show that the cladding material may serve dual purposes since it can also be applied as a means to efficiently outcouple single photons from the nanophotonic waveguide into an optical fiber. Our proposal paves the way for a highly efficient fiber-coupled source of indistinguishable single photons based on a planar nanophotonic platform.
\end{abstract}

%
%
%
%
%

\section{Introduction and motivation}
Research on self-assembled quantum dots (QDs) in photonic nanostructures has witnessed significant progress within the last decade, and the operational principles of highly coherent and efficient photon-emitter coupling have been demonstrated on various platforms~\cite{Lodahl2015}. Consequently, it is now timely to engineer and design efficient and robust quantum-photonic devices and assess the practical limits of these. One enabling device is a deterministic and coherent single-photon source~\cite{Lounis2005} with potential applications spanning the range from dedicated quantum simulators~\cite{Aspuru-Guzik2012} over photonic quantum networks~\cite{Kimble2008,Lodahl2017} to photonic quantum computing~\cite{OBrien2010}. These applications generally require highly efficient, pure, and indistinguishable single-photon sources preferentially implemented in a scalable solid-state device. As the list of candidates for solid-state based sources keeps growing~\cite{Aharonovich2016}, the InGaAs/GaAs QD-based sources have shown great performance for each of these criteria~\cite{He2013,Arcari2014,Sapienza2015,Somaschi2016,Ding2016,Daveau2016,Kirsanske2017}. These sources have the advantage of being easily integrated in nanophotonic structures~\cite{Lodahl2015}, which may enable the on-chip integration with complex functionalities~\cite{OBrien2010}.

The progress in single-photon sources based on QDs has been largely a consequence of the thorough understanding of the coherence properties. Three main decoherence processes have been identified that are relevant in the case of resonant excitation of QDs: charge fluctuations in the electrostatic surrounding of the QD~\cite{Kuhlmann2013}, spin noise arising from the coupling to the nuclear spin bath~\cite{Urbaszek2013}, and the coupling to acoustic phonons~\cite{Besombes2001}. These processes all act on different time scales. Importantly, the first two are slow processes where the environment changes on a long time scale (up to $\sim\si{\micro\second}$) and can be efficiently reduced in electrically contacted structures, which has led to the demonstration of near-transform-limited emission lines~\cite{Kuhlmann2015,Lobl2017}. Very recently such performance was also achieved in nanophotonic waveguides~\cite{Thyrrestrup2018}, which paves the way for a highly efficient and coherent planar photon-emitter interface.
In contrast, the phonon dephasing process is fast ($\sim\si{\pico\second} - \si{\nano\second}$), which is of the order of or faster than the time scale of the QD spontaneous emission. The phonon dephasing constitutes the fundamental limit to the degree of indistinguishability (ID) of single-photon emission from a QD, which is typically measured by interfering two subsequently emitted photons from the QD and is therefore unaffected by slow processes. 

The phonon decoherence can be identified in the emission spectrum of a resonantly excited QD as a broadening of the narrow emission peak (zero-phonon line, ZPL) as well as broad sideband peaks~\cite{Besombes2001,Borri2001,Bayer2002}.
 For QDs in a homogeneous material (bulk), the phonon sideband and the broadening of the ZPL can be explained by a linear (absorption and emission of phonons) and quadratic exciton-phonon coupling (elastic scattering processes), respectively~\cite{Besombes2001,Krummheuer2002,Mahan,Muljarov2004}. In photonic nanostructures, the picture is more involved and additional linear contributions to the broadening of the ZPL have been identified~\cite{Lindwall2007,Tighineanu2018}, which constitute the dominant dephasing mechanism at low temperatures. The sideband photons may readily and efficiently be removed by spectral filtering. An example is filtering with a cavity, which enhances the emission through the ZPL, in which case the overall efficiency of the source is barely reduced~\cite{Iles-Smith2017a}. Importantly and as a limiting upper bound, sideband emission amounts typically to only $\sim 10 \%$ of the total emission for a QD in a bulk sample at the relevant operation temperature of about \SI{4}{\kelvin}~\cite{Lodahl2015} and for current devices this is not the main limitation for the efficiency. As a consequence, the broadening of the ZPL is the fundamental decoherence process limiting the ID of QD single-photon sources.

\begin{figure}
\centering
\includegraphics{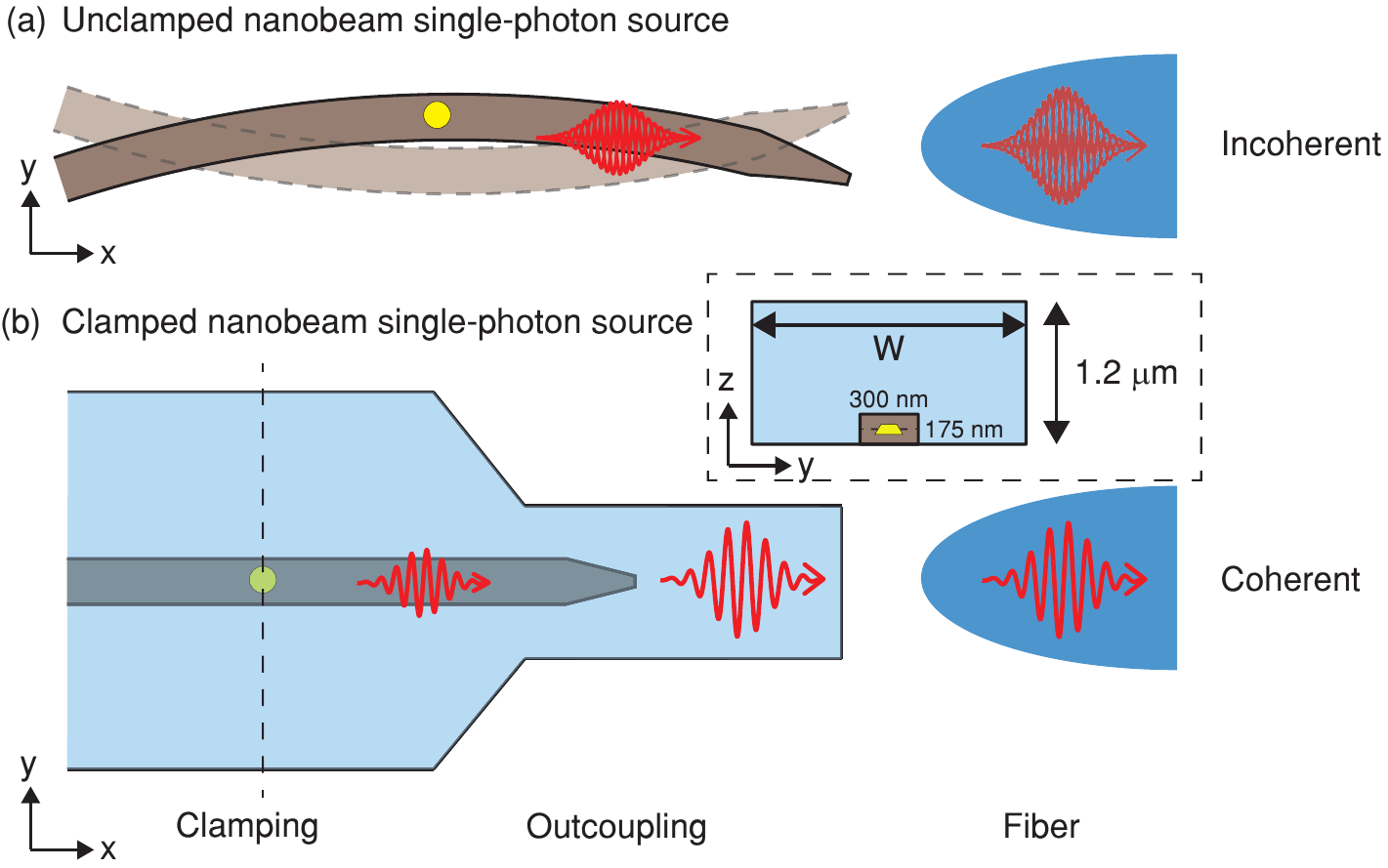}
\caption{Schematics of the mechanical clamping of a suspended nanobeam waveguide. (a) Top view of a regular GaAs waveguide (grey) with a tapered end to couple photons to a single-mode lensed fiber (dark blue). Interactions of the QD (yellow circle) with phonons (the F(1,1)$_\mathrm{y}$ mode is shown here) lead to decoherence and reduce indistinguishability of photons. (b) A layer of a low-index material (light blue) damps fundamental vibrational modes to increase the indistinguishability (ID). Additionally, the cladding material is used as an outcoupling waveguide that can be properly modematched to a lensed fiber. The inset shows the cross-section of the device as indicated by the dashed line in the clamping region.}
\label{fig:fig1}
\end{figure}

The aim of this article is to investigate the theoretical limits to the ID and efficiency of a nanophotonic single-photon source based on a QD embedded in a single-photonic-mode suspended GaAs nanobeam waveguide, c.f. Figure~\ref{fig:fig1}(a). Due to the one-dimensional  geometry, phonon decoherence processes are stronger than in bulk systems because of the enhanced phonon density of states at low frequencies~\cite{Lindwall2007,Tighineanu2018,Massimo1996,Galland2008}.
We propose a realistic device that overcomes these limitations by suppressing the relevant phonon modes with a low refractive index cladding material that effectively clamps the system while preserving efficient optical coupling of the QD to the GaAs waveguide, c.f. Figure~\ref{fig:fig1}(b). We furthermore show that the cladding material can serve the additional purpose of coupling the collected single photons with high efficiency into a single-mode optical fiber. The proposal shows a clear pathway to a fiber coupled highly efficient source of indistinguishable photons.

\section{Theory of the exciton-phonon interaction and numerical modelling}

The calculation of the ID follows the derivations of Reference~\cite{Tighineanu2018} and the key results are summarized below. The interaction between a QD exciton and an acoustic phonon mode is well described by the electron-phonon deformation-potential interaction~\cite{Besombes2001,Muljarov2004,Tighineanu2018,Takaghara1999}, which characterizes the effects on the bandstructure due to strain. We assume that at low temperatures only the exciton ground state $\ket{\psi_1}$ is populated. The interaction potential can be split into a linear part (emission or absorption of a phonon) and a quadratic part (elastic scattering of a phonon). After a delta-pulse excitation at $t=0$, the coherence decay can be described by:
\numparts
\begin{eqnarray} \label{coherence}
P(t)&=\text{exp}(-i\mu_F(t)+K_L(t)+K_Q(t)),\\[4pt]
K_L(t)&=-\frac{1}{2\hbar^2}\sum_{\bi{q}}|L_{\bi{q}}^{11}|^2\left[(2N_{\bi{q}}+1)\left(\frac{\text{sin } \frac{\omega_{\bi{q}} t}{2}}{\frac{\omega_{\bi{q}}}{2}}\right)^2+\frac{2i}{\omega_\bi{q}^2}(\text{sin } \omega_\bi{q} t -\omega_{\bi{q}} t)\right],\label{K_L}\\[4pt]
L_{\bi{q}}^{11} &= M_{\bi{q}e}^{11}-M_{\bi{q}h}^{11},\qquad M_{\bi{q}a}^{11} = D_{a} \bra{\psi_1}\nabla\cdot{\bi{u}_{\bi{q}}(\bi{r_{a}})}\ket{\psi_{1}},\label{Matrixelement}\\[8pt]
K_Q(t)&=-\Gamma_{3D}t,
\end{eqnarray}
\endnumparts
where $\omega_{\bi{q}}$ is the frequency of the phonon mode $q$ with corresponding volumetric strain $\nabla\cdot\bi{u}_{\bi{q}}(\bi{r_{a}})$, evaluated at the position of the electron $\bi{r_{e}}$ and the hole $\bi{r_{h}}$. The vector $\bi{q}$ denotes both the phonon mode and its momentum. $D_{a}$ is the deformation potential for electrons ($a=e$) and holes ($a=h$), and $N_{\bi{q}}$ represents the Bose-Einstein distribution. We consider a spherical QD with a parabolic confinement potential and use for the calculations the ground and first excited single-particle states in the absence of magnetic fields or spin-flip processes. Details about the derivation as well as the expression for the dephasing rate of a bulk system $\Gamma_{3D}$ can be found in Reference~\cite{Tighineanu2018} and the used parameters are summarized in~\cite{parameters}. The term $i\mu_F(t)$ has no effect on the coherence of the system and leads to a spectral shift.

As aforementioned, the interaction with phonons leads to a broad sideband as well as a narrow peak in the QD emission spectrum, see red curve in Figure \ref{fig:fig5}(c). The phonon sideband originates from the emission or absorption of high-energy phonons at the short time scale of a few picoseconds (large $\omega_{\bi{q}}$ contributions to $K_L$). In experiments, the phonon sideband can be efficiently filtered out to reach a high ID~\cite{He2013,Ding2016,Kirsanske2017}, cf. green curve in Figure \ref{fig:fig5}(c). Therefore, we do not consider the phonon sideband for the ID calculation but focus on the broadening of the ZPL, which is the fundamental limitation to the ID of photons. The width of the ZPL is determined by slower interaction processes compared to the sideband. These include second-order photon scattering processes ($K_Q$) and the emission and absorption of low-energy phonons (small $\omega_{\bi{q}}$ contributions to $K_L$). Since the phonon density of states in a bulk material is proportional to $\omega_{\bi{q}}^2$ and vanishes for low energies, $K_L$ does in this case not contribute to the broadening of the ZPL. $K_Q$ is to a very good approximation independent of the photonic nanostructure for realistic structures that are larger than several tens of nanometers. This is because its dominating contribution stems from phonon modes with a wavelength comparable to the size of the QD. Such waves are much smaller than the spatial scale of the nanostructure and hence a bulk description approximates $K_Q$ well~\cite{Tighineanu2018}. However, the modified phonon density of states in, e.\,g., waveguides and membranes (one- and two-dimensional photonic nanostructures) implies that low-energy phonons broaden the ZPL additionally and it turns out that these are dominant at low temperatures in comparison to the quadratic coupling. It is the contribution of these phonon modes that is investigated in the present article. Thereby, the decisive parameters  of the phonon modes are the frequency ($\omega_{\bi{q}}$) and the volumetric strain ($\nabla\cdot\bi{u}_{\bi{q}}$) as can be seen from Eqs. (\ref{K_L}) and (\ref{Matrixelement}). A mode with high volumetric strain and low frequency over a large window of wavenumbers will dephase a QD significantly.

The ID of the emitted photons is experimentally recorded by measuring the quantum interference between two photons in a Hong-Ou-Mandel experiment. The ID ranges from 0 (distinguishable photons) to 1 (fully indistinguishable photons) and can be expressed in the frequency domain as~\cite{Iles-Smith2017a}
\begin{eqnarray} \label{TPI}
\text{ID} = \frac{\int\!\!\int_{-\infty}^\infty \text{d}\omega_1 \text{d}\omega_2 \hspace{3pt} |\langle \sigma_-(\omega_1)\sigma_+(\omega_2)\rangle|^2 \hspace{3pt} |h(\omega_1)|^2 \hspace{3pt} |h(\omega_2)|^2}{\left[\int_{-\infty}^\infty \text{d}\omega \hspace{3pt} \langle \sigma_-(\omega)\sigma_+(\omega)\rangle \hspace{3pt} |h(\omega)|^2\right]^2},
\end{eqnarray}
where $ h(\omega)$ is a spectral filter that can be implemented in the experiment. $\langle \sigma_-(\omega_1)\sigma_+(\omega_2)\rangle$ is the two-color spectrum expressed by the raising and lowering operators of the QD. Spontaneous radiative emission is included as a phenomenological decay yielding the following expression for the two-color spectrum
\begin{eqnarray} \label{spectrum}
\langle \sigma_-(\omega_1)\sigma_+(\omega_2)\rangle = \int\!\!\!\int_0^{\infty} \text{d}t \text{d}\tau \hspace{3pt} P(|t-\tau|)\text{e}^{-\Gamma_\text{rad}(t+\tau)/2} \hspace{4pt} \text{e}^{i\omega_2\tau-i\omega_1 t},
\end{eqnarray}
where $\Gamma_\text{rad}$ is the radiative decay rate. This expression holds in the weak-coupling (Purcell) regime of light-matter interaction suitable for waveguides and weak cavities.

\begin{figure}
\subfloat{
\includegraphics[width=.5\textwidth]{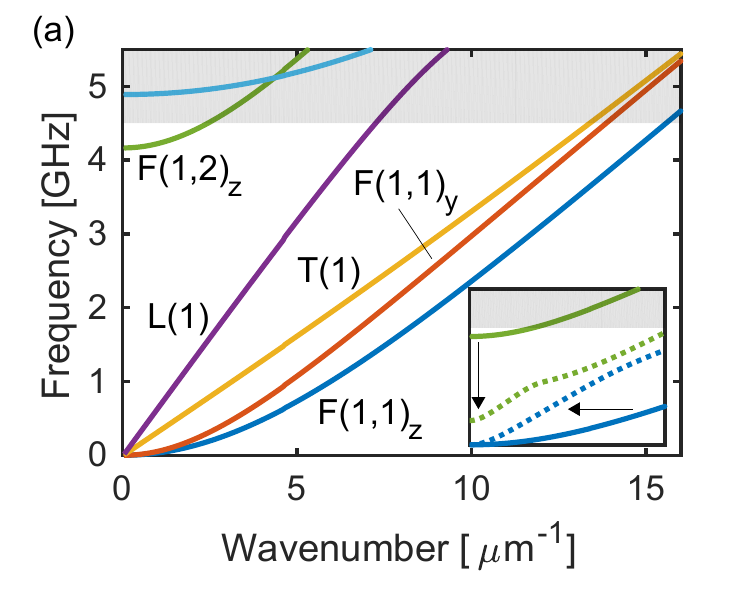}}
\subfloat{
\includegraphics[width=.5\textwidth]{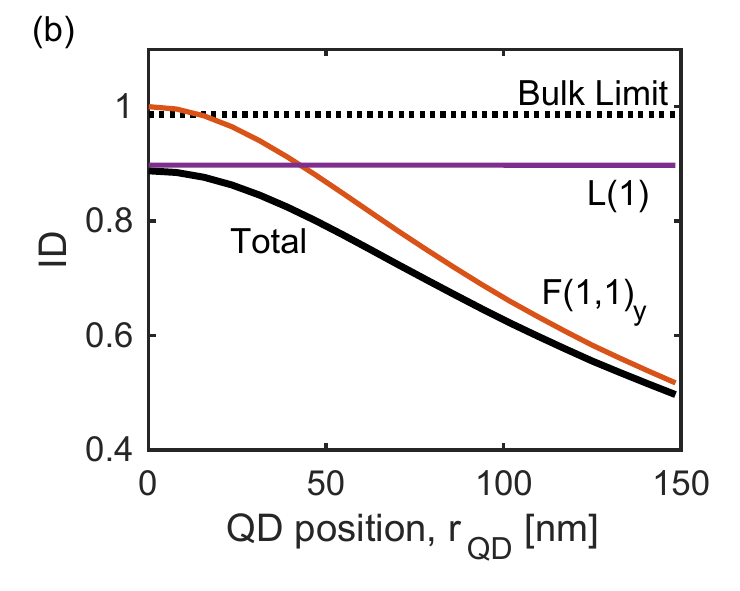}}\\
\subfloat{
\centering
\includegraphics[width=\textwidth]{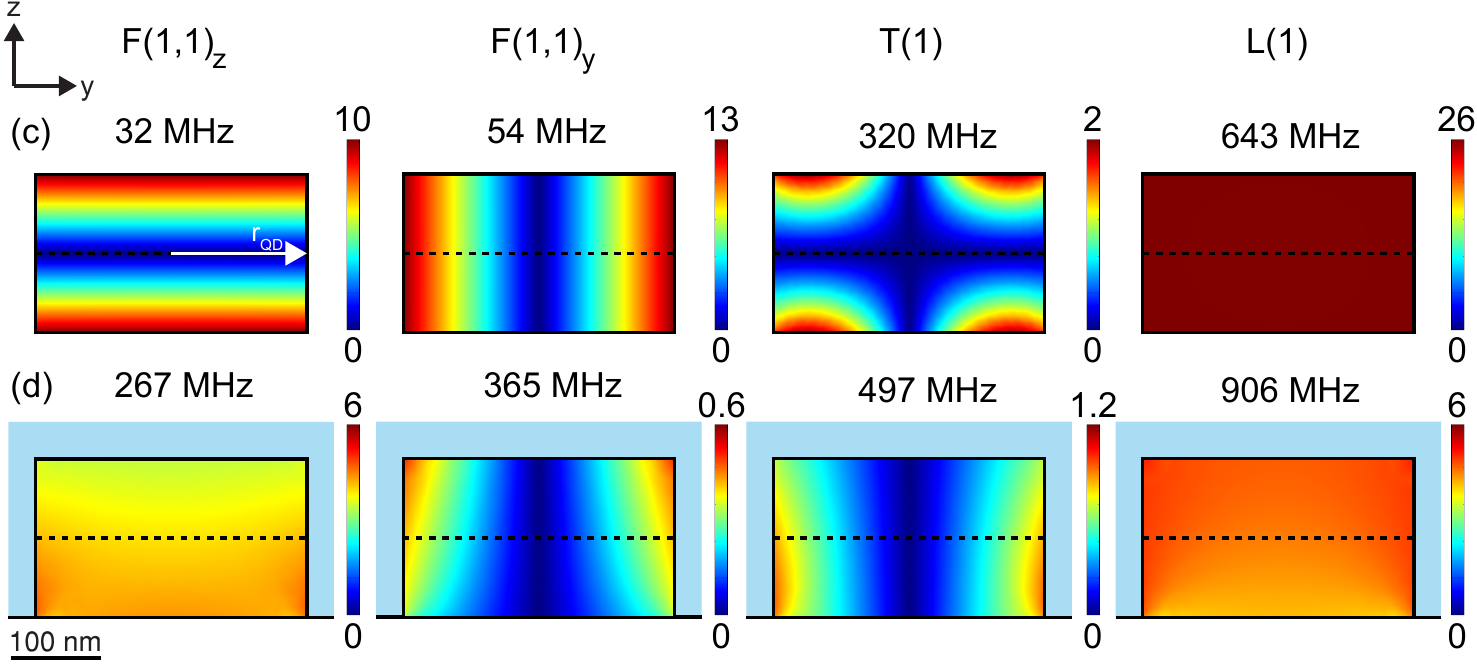}}
\caption{(a) Dispersion relation of the first six modes of an unclamped nanobeam waveguide. The detailed description is given in the main text. The inset shows how the dispersion of the modes F(1,1)$_\mathrm{z}$ and F(1,2)$_\mathrm{z}$ changes as we go from an unclamped waveguide (solid lines) to a clamped waveguide with $W_{\mathrm{Cladding}}=\SI{2}{\micro\meter}$ (dotted lines). (b) ID as a function of QD location along the dashed line in (c) going from center to the edge of the nanobeam waveguide. The black curve represents the total ID, including the quadratic coupling. The red and the purple curve represent the contribution from the F(1,1)$_\mathrm{y}$ and L(1) modes, respectively (color coding as in (a)). All other modes have negligible contribution to the ID. The black dotted line shows the contribution from the quadratic coupling (bulk limit). Note that the individual contributions are not directly additive. T = \SI{4}{\kelvin}. (c),~(d) Absolute magnitude of the volumetric-strain profiles ($|\nabla\cdot\bi{u}_{\bi{q}}|$) of the fundamental phonon modes in the cross-sections of an unclamped (c) and a clamped (d) nanobeam waveguide with $W_{\mathrm{Cladding}}=\SI{2}{\micro\meter}$. The cladding material in (d) is indicated by a light blue color, and the corresponding strain profiles of this material have been left out. The dashed horizontal lines indicate the investigated QD positions. All profiles were calculated for the same wavenumber, $q = \SI{1}{\micro\meter^{-1}}$, and the corresponding frequencies are noted above the structure. The strain scales are in units of $10^{-13}$ fractional change in volume.}
\label{fig:fig1.5}
\end{figure}

It is possible to calculate the ID analytically for a QD located in the center of highly symmetric structures, for example a membrane (two dimensions) or a cylinder (one dimension) where the mechanical modes are well known~\cite{Lindwall2007,Tighineanu2018}. However, for experimentally relevant photonic nanostructures, numerical evaluation is required in order to understand these more complex devices in particular regarding the dependence on the spatial position of the QD. In the present work we use three-dimensional finite-element analysis to simulate the phonon modes of suspended nanobeam waveguides.
We model first an unclamped nanobeam waveguide consisting of GaAs~\cite{parameters} surrounded by air with dimensions suitable for experimentally relevant efficient single-photon sources, i.\,e., \SI{300}{\nano\meter} width and \SI{175}{\nano\meter} height~\cite{Kirsanske2017}. For the clamped nanobeam waveguide a layer of cladding material is added on top, as shown in the inset of Figure \ref{fig:fig1}. The use of Floquet boundary conditions at both ends of a slice of the nanobeam enables the simulation of mechanical waves with any given wavenumber $q$ in just a thin portion (\SI{60}{\nano\meter}) of the waveguide. Because of computational limitations, we can only consider a finite number of wavenumbers and modes. As discussed above, low-energy phonons dominate the broadening of the ZPL through $K_L$. For unclamped nanobeam waveguides, only the fundamental mechanical modes contribute significantly because the higher-order modes are much higher in energy than the natural linewidth of the QD and can be readily filtered out. An increase of the cross-section of the structure (due to the cladding material) shifts down the cut-off frequencies of the higher-order modes (see green lines in the inset of Figure \ref{fig:fig1.5}(a)), which results in the contribution of these modes to the dephasing since they overlap with the ZPL and cannot be filtered out. However, the dephasing contribution from each individual mode is significantly reduced. In the limit of the cross-sectional area becoming infinite, the energy density around $\omega \sim 0$ tends to zero and the bulk behavior is recovered in which the linear interaction plays no role. Therefore, the net result is that the overall coherence is enhanced when adding a sufficient amount of cladding material. We restrict the number of modes by solving in a frequency window from 0 to \SI{4.5}{\giga\hertz}. This window is large enough to include the main contribution to the dephasing. For each mode at a given $q$, the frequency $\omega_{\bi{q}}$ and the normalized volumetric strain $\nabla\cdot{\bi{u}_{\bi{q}}}(\bi{r})$ at the position $\bi{r}$ of the QD can be extracted. We assume a point-like QD because low-energy phonons have a wavelength much longer than the size of the QD ($\bi{r_{e}}=\bi{r_{h}}=\bi{r}$). The computation is repeated for increasing $q$ until no additional modes are found in the \SI{4.5}{\giga\hertz} window any more. The extracted values are summed according to Eq. (\ref{K_L}) and the coherence $P(t)$ is calculated. We assume a Lorentzian spectral filter $|h(\omega)|^2 = \Gamma_\text{f}^2/(\omega^2+\Gamma_\text{f}^2)$ with half-width-at-half-maximum of $\Gamma_\text{f}=\SI{1.5}{\giga\hertz}$ and calculate the ID as given by Eqs.~(\ref{TPI}) and~(\ref{spectrum}).

\section{Phonon modes in an unclamped nanobeam waveguide and single-photon indistinguishability}
The nanobeam waveguide exhibits four fundamental vibrational modes that approach $\omega=0$ for small wavenumbers: two flexural F(1,1)$_\mathrm{y}$ and F(1,1)$_\mathrm{z}$, one torsional T(1), and one longitudinal L(1) mode~\cite{Mason1966}, c.f. Figure \ref{fig:fig1.5}(a) for the dispersion relation and Figure \ref{fig:fig1.5}(c) for the volumetric-strain profiles. The volumetric strain of a phonon characterizes the local periodic change in volume, which leads to a change in the band structure of a QD and therefore to a time-dependent phase, i.\,e., dephasing. During the growth of self-assembled QDs, their vertical position is very well controlled (i.\,e., with atomic mono-layer precision) while the lateral position is generally not controlled unless advanced positioning methods are implemented, which adds complexity to the device fabrication. It is therefore important to study the lateral position dependence of the decoherence processes in order to combat it. The strain profiles in Figure \ref{fig:fig1.5} at a given wavenumber provide insight into the spatial contribution of the volumetric strain of the modes to dephasing. One must keep in mind however, that the weight of the various modes also depends on their mode energy density and the integration over wavenumbers. In the unclamped nanobeam, F(1,1)$_\mathrm{z}$ and T(1) have symmetry planes in the vertical centre of the nanostructure and therefore do not dephase QDs located in this plane since $\nabla\cdot\bi{u}_{\bi{q}}=0$, see Figure \ref{fig:fig1.5}(c). F(1,1)$_\mathrm{y}$ has a symmetry plane in the horizontal centre of the structure. The mode therefore does not contribute to the decoherence of QDs placed in the center but its contribution increases towards the edge. In contrast, L(1) is position independent and so is its contribution to the dephasing. Accordingly, L(1) is found to be the dominant source of decoherence in the center of the waveguide and F(1,1)$_\mathrm{y}$ at the edge, see Figure \ref{fig:fig1.5}(b). We have also evaluated the contributions from higher-order modes and found that they are at least three orders of magnitude weaker than the contributions from F(1,1)$_\mathrm{y}$ and L(1). Note that the individual contributions are not additive to recover the total ID. The analysis shows that, for a typical nanobeam waveguide, the total ID is at best  $\SI{89}{\percent} $ at T = \SI{4}{\kelvin}, assuming a typical QD decay rate in a nanobeam waveguide of $\Gamma_{rad}= \SI{1}{\per\nano\second}$. One approach to improve the ID is to increase the radiative decay rate, which can be done either by Purcell enhancement in a cavity or slow-light waveguide or by increasing the oscillator strength of the QD. In the following section we pursue an alternative approach where clamping of the relevant phonon modes is found to increase the ID significantly.

\section{Phonon clamping of a nanobeam waveguide for improved photon indistinguishability }
In order to improve the ID of the emitted photons, it is necessary to reduce the exciton-phonon coupling at low frequencies that are of the order of the natural linewidth. Adding an auxiliary cladding material on top of the nanostructure, as illustrated in Figure \ref{fig:fig1}(b), will damp the dephasing effect of the fundamental phonon modes. The strain profiles of the whole clamped structure (not shown) are similar to the profiles of the unclamped waveguide (Figure \ref{fig:fig1.5}(c)), i.\,e., one can identify the corresponding modes but with decreased amplitude and modified dispersion. 

The change in dispersion is mode-dependent and determined by the alteration of the material properties as well as the cross-section. As an example, we plotted the dispersion of the mode F(1,1)$_\mathrm{z}$ and its second-order mode F(1,2)$_\mathrm{z}$ in an unclamped and a clamped waveguide in the inset of Figure \ref{fig:fig1.5}(a). The former mode is squeezed towards lower wavenumbers due to the clamping, which results in a higher energy at a given wavenumber for the fundamental modes, compare also the frequencies in Figure \ref{fig:fig1.5}(c) and (d). However, the cut-off frequency of the higher-order mode is shifted down in energy. These two effects on the dispersion relation lead to a reduced influence of the fundamental modes to the decoherence and to the need of considering higher-order modes for a correct ID calculation. Other effects of the clamping are the decrease of the density of states at low energies and altered volumetric-strain profiles, resulting in modified volumetric-strain amplitudes, see Figure \ref{fig:fig1.5}(d). Note that the influence of F(1,1)$_\mathrm{z}$ to the decoherence is now increased since the QD is no longer at the symmetry point where the strain vanishes. Nevertheless, the damping of the other modes outweigh this increase, so that the overall effect is still an increase of ID. F(1,1)$_\mathrm{y}$ experiences the strongest suppression, which leads to a decrease of the QD-position dependence on the dephasing. The influence of L(1) is damped significantly as well because of a higher energy and a lower amplitude in comparison to the unclamped case. Regarding the choice of cladding material, the simulations showed that a high Young modulus, a low density, and a Poisson ratio close to 0.5 are beneficial to suppress phonons. Importantly, the cladding material must have a low refractive index to keep the optical mode well localized within the GaAs for a good overlap with the QD. We choose SiO$_2$ as cladding in the simulations~\cite{parameters}, which is also a suitable material to be used an outcoupling waveguide for the photons.

\begin{figure}
\centering\includegraphics{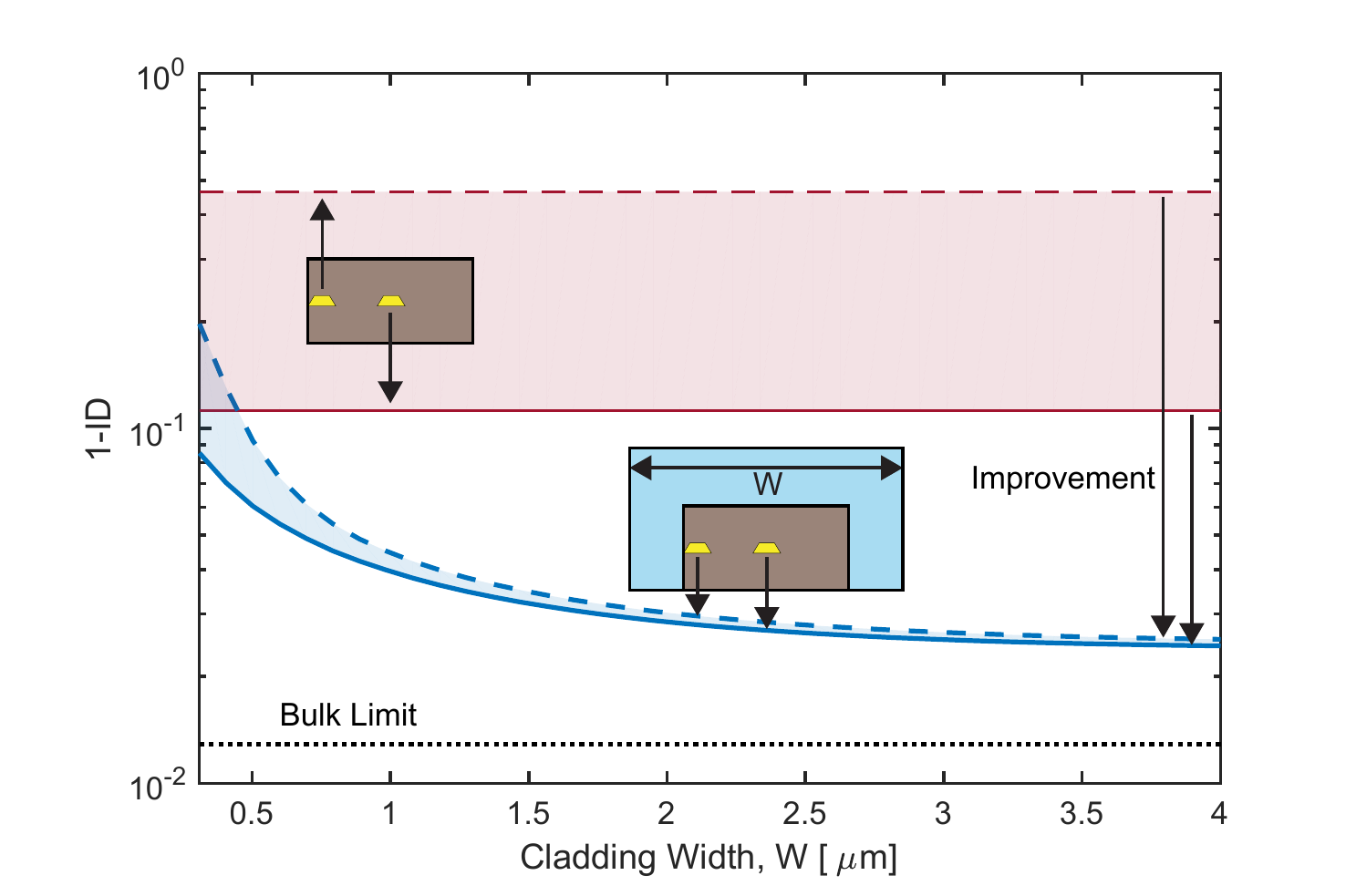}
\caption{The deviation from unity ID in a clamped structure as a function of the cladding width is plotted as the blue curves with the corresponding case of unclamped waveguides shown by the red curves. The solid (dash-dotted) lines represent a QD positioned at the center (edge) of the GaAs nanobeam (see inset) and the filled regions cover the QD positions in between. The height of the SiO$_2$ cladding is fixed to \SI{1.2}{\micro\meter}. The limit for a QD in a bulk medium of ID $\sim 99 \%$ is displayed as black dotted line. $T = \SI{4}{\kelvin}$.}
\label{fig:fig2}
\end{figure}

The effect of clamping on the ID can be seen in Figure \ref{fig:fig2}. The improvement in coherence scales with the size of the cladding material. The height of the cladding material is fixed to $H_{\mathrm{Cladding}}=\SI{1.2}{\micro\meter}$ to match the conditions for efficient outcoupling (see below), while the width can be changed freely. The ID increases drastically with increasing cladding width until it converges to the limit of an infinite membrane of the same height, close to the fundamental bulk limit. This improvement happens already for cladding structures of a few \si{\micro\meter} width. Remarkably, the position dependence of the ID vanishes almost completely. As an example, at $W_{\mathrm{Cladding}}=\SI{4}{\micro\meter}$ the ID is around \SI{97.6}{\percent} both in the center and at the edge. The corresponding limit for a QD in a bulk medium is  ID $=\SI{98.7}{\percent}$.

\begin{figure}
\subfloat{
\includegraphics[width=.5\textwidth]{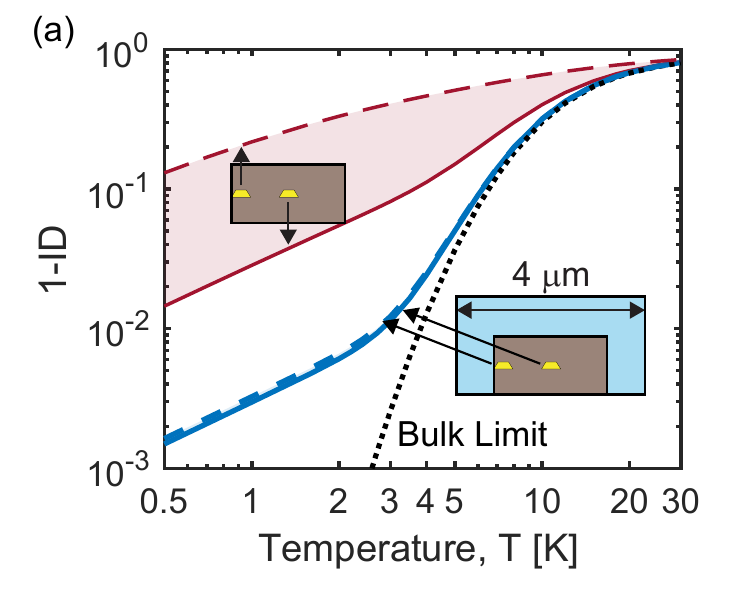}
}
\subfloat{
\includegraphics[width=.5\textwidth]{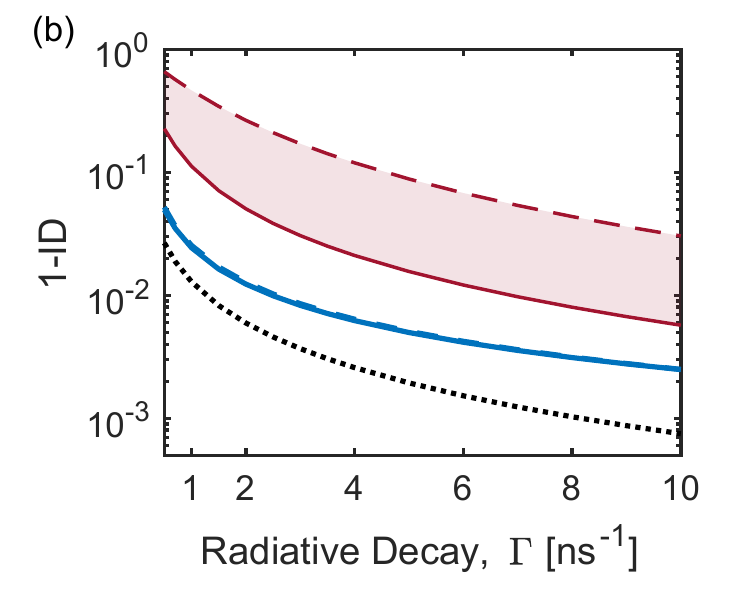}
}
\caption{Deviation from perfect ID depending on (a) the temperature, $T$, and (b) the radiative decay rate, $\Gamma_{rad}$, of the QD. The blue lines represent a clamped nanobeam with $W_{\mathrm{Cladding}}=\SI{4}{\micro\meter}$ while the red lines correspond to the unclamped nanobeam. The solid (dash-dotted) lines represent a QD at the center (edge) of the GaAs nanobeam. The bulk limit is given by the black dotted line. (a) $\Gamma_{rad} = \SI{1}{\nano\second^{-1}}$. (b) $T = \SI{4}{\kelvin}$.}
\label{fig:fig4}
\end{figure}

To reach ID $> \SI{99}{\percent}$, it is essential to improve the bulk limit. One possibility is to minimize the operational temperature of the device, which directly lowers the phonon occupation number. Figure \ref{fig:fig4}(a) shows the temperature dependence of the ID for QDs in an unclamped nanobeam waveguide, in a clamped nanobeam with $W_{\mathrm{Cladding}}=\SI{4}{\micro\meter}$, and in a bulk medium. At high temperatures, the dephasing is dictated by the quadratic coupling term which is a two-phonon process and therefore scales stronger with the temperature than the linear coupling ($K_Q \propto T^{11}$ at low $T$, $K_Q \propto T^{2}$ at high $T$, $K_L \propto T$ at all $T$)~\cite{Tighineanu2018}. At lower temperatures, the linear coupling prevails in one-dimensional structures. The temperature at which this transition occurs depends on the cross-section of the structure ($\sim 10 - \SI{20}{\kelvin}$ for the unclamped case, $\sim \SI{4}{\kelvin}$ for the clamped case). Due to the suppression of the linear coupling by clamping we can achieve almost unity ID in clamped structures at temperatures below \SI{4}{\kelvin}. 
For example,  ID $= \SI{99.5}{\percent}$ can be reached at a temperature of 1.6K, which can be achieved with standard closed-cycle cryostats~\cite{Kirsanske2017,Thyrrestrup2018}.

Improving the radiative decay rate of the QD can also lead to a higher ID. This can be done either by having a large QD oscillator strength or by Purcell enhancement in, e.\,g., a cavity or a slow-light waveguide. As used before, a typical radiative decay rate of InAs/GaAs QDs in a bulk medium is $\Gamma_{\mathrm{rad}}=\SI{1}{\nano\second^{-1}}$~\cite{Lodahl2015}. Figure \ref{fig:fig4}(b) shows that the ID of faster-decaying QDs is increasing drastically. The ID is very responsive to changes in $\Gamma_{\mathrm{rad}}$ in the region of low enhancement and an ID value over \SI{99}{\percent} can be reached already with $\Gamma_{\mathrm{rad}}=\SI{3}{\nano\second^{-1}}$ with a cladding layer of $\SI{4}{\micro\meter}$ at $T = \SI{4}{\kelvin}$. Even a QD in an unclamped waveguide can achieve high ID, however in this case it is highly sensitive to the QD position in the waveguide. Therefore, a combination of several methods (clamping, low $T$, high $\Gamma_{\mathrm{rad}}$) could be deployed to reach a high yield of efficient coherent single-photon sources.

\section{Efficiency of single-photon sources with cladding}
An ideal single-photon source is capable of providing a large number of mutually indistinguishable single photons. An important figure-of-merit is the overall efficiency of the source ($\eta_{\mathrm{total}}$). It is defined as the probability that an excited QD leads to a photon exiting the chip and is coupled to a usable optical mode. The total efficiency comprises several intermediate efficiencies, $\eta_{\mathrm{total}}=\beta \hspace{4pt} \eta_{\mathrm{outcoupling}} \hspace{4pt} \eta_{\mathrm{filtering}}$. Firstly, the excited QD must emit the photon into the desired optical mode that is supported by the waveguide, whose probability is the spontaneous emission $\beta$-factor. Secondly, the photon in the waveguide must be coupled out from the chip ($\eta_{\mathrm{outcoupling}}$) and, lastly, a filter has to be implemented in order to remove the phonon sidebands ($\eta_{\mathrm{filtering}}$). In the following we evaluate the efficiency of a QD single-photon source in a clamped nanobeam waveguide.

\begin{figure}
\subfloat{
\includegraphics[width=.5\textwidth]{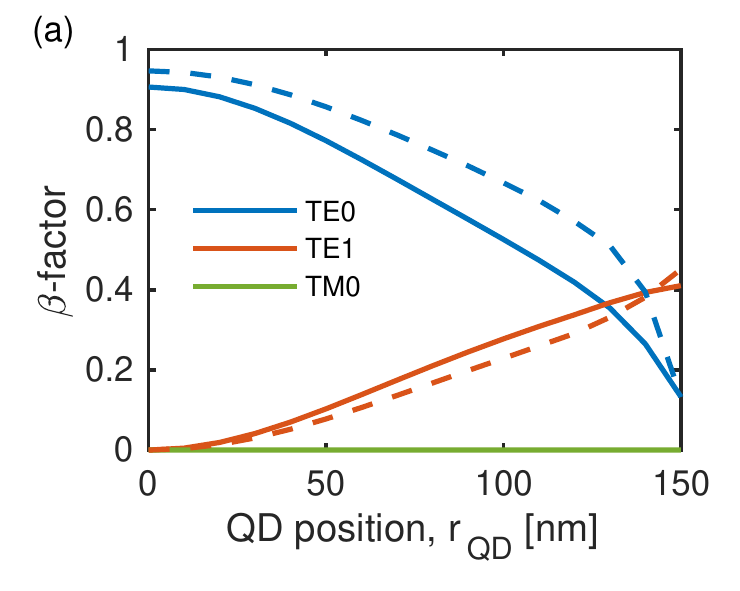}}
\subfloat{
\includegraphics[width=.5\textwidth]{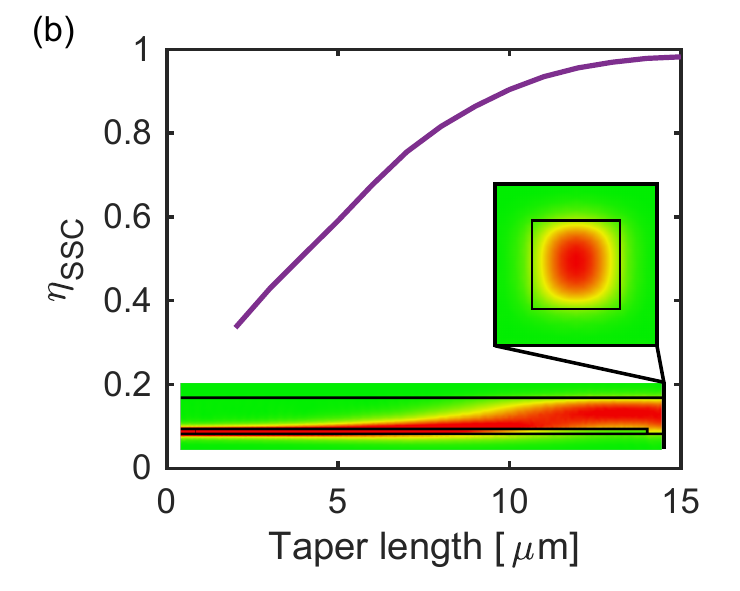}}\\
\centering
\subfloat{
\includegraphics[width=.5\textwidth]{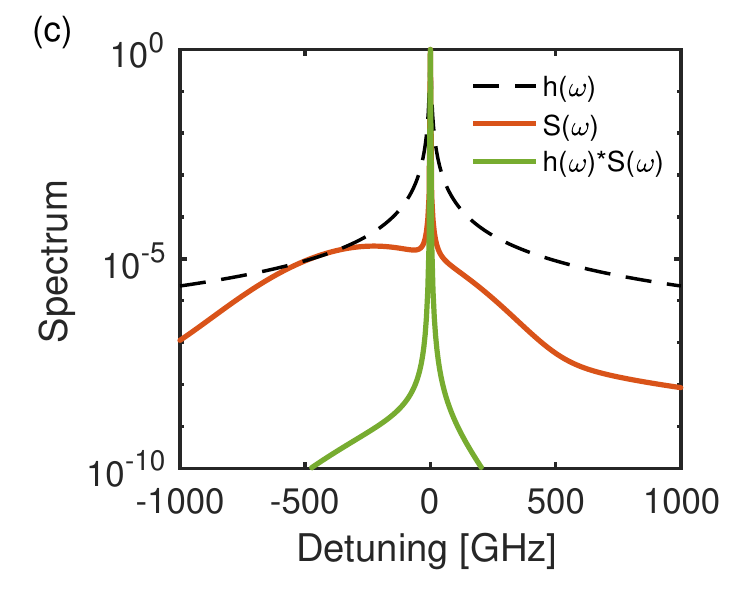}}
\caption{Evaluation of the three contributions to the total source efficiency. (a) The $\beta$-factor as a function of the QD position from the center of the GaAs waveguide. The dashed lines represent an unclamped nanobeam and the solid lines represent a nanobeam waveguide with cladding. (b) The mode conversion efficiency $\eta_{\mathrm{SSC}}$ depending on the taper length in an inverted taper design as indicated in Figure \ref{fig:fig1}(b). The width of the taper tip is \SI{60}{\nano\meter} and the cross-section of the cladding material is \SI{1.2}{\micro\meter} x \SI{1.2}{\micro\meter}. The insets show how the TE$_0$ mode transfers from the GaAs into the cladding material and the mode profile in the cladding after tapering out to a taper with length of \SI{15}{\micro\meter}. (c) The emission spectrum $S(\omega)$ of a QD (red curve), which can be efficiently filtered by a Lorentzian filter $h(\omega)$ with $\Gamma_\text{f}=\SI{1.5}{\giga\hertz}$ (dashed black line) to achieve a highly coherent output spectrum from the ZPL $h(\omega)S(\omega)$ (green curve). $T = \SI{4}{\kelvin}$.}
\label{fig:fig5}
\end{figure}

It is desirable that the QD radiation couples efficiently to only one optical mode, i.\,e., the fundamental TE mode (TE$_0$), to achieve both high beta-factor and large out-coupling efficiency into a fiber. This can be achieved by reducing the number of modes supported by the waveguide and suppressing coupling of QD emission to unwanted waveguide modes. The $\beta$-factor of QD emission into a given optical waveguide mode can be calculated with three-dimensional finite-element simulations by placing a dipole in the waveguide and evaluating the fraction of power coupled into this mode. The calculation was done for both the unclamped and the clamped waveguide and is shown in Figure \ref{fig:fig5}(a) as a function of the QD position. We study a dipole orientation along the y-direction which maximises the coupling to the TE$_0$ mode compared to the one along the x-direction~\cite{Kirsanske2017}. The simulation for the clamped waveguide is valid for all cladding widths above \SI{1}{\micro\meter} since the optical modes barely change for thicker layers. In an unclamped waveguide, an in-plane dipole sitting in the vertical mirror plane of the waveguide only couples to TE modes. Here, the dipole mainly couples to two modes, TE$_0$ and TE$_1$, where TE$_0$ is the desired symmetric mode. In the center of the nanobeam cross-section, the coupling efficient into the TE$_0$ mode is maximized and the coupling into the TE$_1$ mode vanishes due to its antisymmetric y-polarized electric field profile. This leads to a high $\beta$-factor of \SI{95}{\percent} for the TE$_0$ mode. The remaining \SI{5}{\percent} represent coupling to the non-guided radiation continuum. At the edge of the waveguide, the overlap of a y-dipole with the TE$_0$ mode decreases while the overlap with the TE$_1$ mode increases, leading to a reduction of the $\beta$-factor into the TE$_0$ mode to \SI{14}{\percent}. Adding a cladding material on top of the waveguide alters the $\beta$-factor since the change of refractive index in the surrounding of the waveguide modifies the supported optical modes. Due to the asymmetry of the structure, a coupling of a y-dipole into the TM$_0$ mode is expected, which might be a concern for decreasing the $\beta$-factor into the TE$_0$ mode. However, the calculated coupling coefficient into the TM$_0$ mode is negligible small (green line in Figure \ref{fig:fig5}(a)). The profiles of the TE$_0$ and TE$_1$ modes are dragged in the direction of the cladding material due to the higher refractive index compared to air. This leads to a decrease of the $\beta$-factor into the predominant TE$_0$ mode by only a few percent. We obtain $\beta = \SI{91}{\percent}$ into the TE$_0$ mode for a QD placed in the center of the clamped waveguide.

The outcoupling of the generated single photons has been done in the literature by several approaches using either vertical (from the top)~\cite{Luxmoore2013,Faraon2008,Javadi2015} or transverse collection (from the side)~\cite{Daveau2016,Laucht2012}. Transverse coupling, for example using evanescent coupling to tapered fibers or edge-coupling with lensed fibers, can be favorable to achieve near-unity collection efficiency, since loss associated with polarization filtering could be avoided. For the GaAs platform, it was shown that an adiabatic tapering of the waveguide (indicated in Figure \ref{fig:fig1}(a)) can lead to an efficient mode transfer of \SI{80}{\percent} into a tapered fiber~\cite{Kirsanske2017}. Edge coupling can be realized using spot-size converters made of inverted tapers, a technique which is widely used in silicon photonics. The highly-confined light in the waveguide is converted into a larger mode in the cladding material to match the optical mode of the fiber (spot-size conversion)~\cite{Fan1999,McNab2003,Roelkens2006}. The cladding material on top of a GaAs waveguide can be designed to form such an inverted taper, as illustrated in Figure \ref{fig:fig1}(b) and the inset of Figure \ref{fig:fig5}(b). The outcoupling efficiency is defined as the product of the efficiency of the spot-size converter ($\eta_{\mathrm{SSC}}$) and reflection losses at the two interfaces (cladding/air $\eta_{\mathrm{Cladding\rightarrow Air}}$ and air/fiber $\eta_{\mathrm{Air\rightarrow Fiber}}$). To match the beam waist of a lensed fiber (here we assume a Gaussian mode field diameter with $1/e^2$ value of \SI{1.2}{\micro\meter}), a mode converter can be designed using a \SI{1.2}{\micro\meter} x \SI{1.2}{\micro\meter} waveguide with a linear waveguide taper. We calculated $\eta_{\mathrm{SSC}}$ using three-dimensional finite-element calculations for different taper lengths and with a tip width of \SI{60}{\nano\meter}, see Figure \ref{fig:fig5}(b). This is achievable with electron beam lithography in current fabrication processes~\cite{Midolo2015a}. For a length of \SI{15}{\micro\meter} a conversion efficiency above \SI{98}{\percent} is reached (see the inset of Figure \ref{fig:fig5}(b)), which can be further improved by narrowing down the tip width. In addition, the reflection at each interface amounts to $\eta_{\mathrm{Cladding\rightarrow Air}} = \eta_{\mathrm{Air\rightarrow Fiber}} \sim\SI{3}{\percent}$, which was calculated using Fresnel equations. This loss can be further minimized by adding anti-reflection coatings, changing the angle of the interface or gluing the fiber directly to the cladding material.

Finally, the phonon sideband has to be filtered out to reach a high ID. To find the efficiency of this process, the QD emission spectrum $S(\omega)=\mathrm{Re}\int_0^{\infty}\mathrm{d}t P(t)\mathrm{exp}(-\Gamma_{\mathrm{rad}}t/2)\mathrm{exp}(-i\omega t)$ is calculated before and after convoluting with a Lorentzian filter, cf.  Figure \ref{fig:fig5}(c). The filtering efficiency ($\eta_{\mathrm{filtering}}$) is given by the ratio of the integrated intensities of the filtered and the unfiltered spectrum. We note that the sidebands originate from short-wavelength phonons, which are negligibly significantly influenced by the extent of the nanostructure. We therefore calculate the sidebands from the results of the bulk system using the same parameters as above~\cite{parameters}. $\eta_{\mathrm{filtering}}$ is temperature-dependent as the phonon occupation decreases with decreasing temperature and the values for the temperatures \SI{1.6}{\kelvin}, \SI{4}{\kelvin} and \SI{10}{\kelvin} can be found in Table \ref{Table1}. The efficiency could be further improved by implementing Purcell enhancement since this increases the fraction of the spectrum emitted into the ZPL.

\begin{table}
\caption{\label{Table1}Efficiencies and ID for a QD in the center of a clamped waveguide for different temperatures. $W_{\mathrm{Clamping}}=\SI{4}{\micro\meter}, W_{\mathrm{Outcoupling}}=\SI{1.2}{\micro\meter}, H_{\mathrm{Cladding}}=\SI{1.2}{\micro\meter}$.}
\begin{indented}
\lineup
\item[]\begin{tabular}{@{}lllllll}
\br
$T$ [\si{\kelvin}] & $\beta$ & $\eta_{\mathrm{outcoupling}}$ & $\eta_{\mathrm{filtering}}$& $\eta_{\mathrm{total}}$ & ID\\
\mr
1.6 & 0.91 & 0.92 & 0.92 & 0.77 & 0.995\\
4 & 0.91 & 0.92 & 0.91 & 0.76 & 0.976\\
10 & 0.91 & 0.92 & 0.72 & 0.60 & 0.682\\
\br
\end{tabular}
\end{indented}
\end{table}

The overall device performance is highly temperature dependent due to the influence of phonons. Indeed both the ID and the overall efficiency depend on temperature due to the contributions from the broadening of the ZPL and the phonon sidebands, respectively. Table 1 lists typical values for experimentally relevant temperatures. At \SI{10}{\kelvin} both ID and efficiency are significantly reduced below unity. However, operating at \SI{4}{\kelvin} or below, we predict an overall efficiency of $\sim\SI{75}{\percent}$ and ID of $>\SI{97.5}{\percent}$ for a QD positioned in the center of the waveguide. This implies that running such a single-photon source at a typical repetition rate of  \SI{80}{\mega\hertz} with resonant pulsed excitation is capable of producing up to 60 million indistinguishable photons per second. Such a source could subsequently be de-multiplexed in order to obtain multiple trains of single photons, which would lead to new opportunities for multi-photon quantum optics experiments \cite{Lodahl2017}.

\section{Conclusion}

We have presented a layout of a novel fiber-coupled QD single-photon source capable of efficiently generating highly indistinguishable photons. The device consists of a high  refractive index nanophotonic waveguide that provides a high $\beta$-factor light-matter interface. The waveguide is furthermore equipped with a lower index cladding material serving the dual purpose of damping the relevant decoherence from phonons and providing an out-coupling waveguide for subsequent efficient coupling to a single-mode fiber. For experimentally feasible temperatures, we predict that the ID of photons can exceed $\SI{99}{\percent}$ and that the cladding material makes ID insensitive to the QD position. Furthermore, we predict a possible total efficiency of $\SI{77}{\percent}$ of collecting the single photons from the QD and subsequently coupling them into a single mode fiber via evanescent transfer to the cladding. This efficiency could readily be improved even further by implementing Purcell enhancement, since it is limited by the contributions from phonon sidebands. The proposal devises a pathway to an integrated and robust on-demand source of coherent photons based on the scalable platform of planar nanophotonic waveguides.

\section*{Acknowledgements}

We gratefully acknowledge financial support from the European Research Council (ERC Advanced Grant "SCALE"), Innovation Fund Denmark (Quantum Innovation Center "Qubiz"), and the Danish Council for Independent Research (Natural Science).


\section*{References}

\begin{thebibliography}{10}

\bibitem{Lodahl2015}
P.~Lodahl, S.~Mahmoodian, and S.~Stobbe, ``{Interfacing single photons and
  single quantum dots with photonic nanostructures},'' {\em Rev. Mod. Phys.},
  vol.~87, no.~2, p.~347, 2015.

\bibitem{Lounis2005}
B.~Lounis and M.~Orrit, ``{Single-photon sources},'' {\em Rep. Prog. Phys.},
  vol.~68, no.~5, p.~1129, 2005.

\bibitem{Aspuru-Guzik2012}
A.~Aspuru-Guzik and P.~Walther, ``{Photonic quantum simulators},'' {\em Nat.
  physics}, vol.~8, no.~4, p.~285, 2012.

\bibitem{Kimble2008}
H.~J. Kimble, ``{The quantum internet},'' {\em Nature}, vol.~453, p.~1023,
  2008.

\bibitem{Lodahl2017}
P.~Lodahl, ``Quantum-dot based photonic quantum networks,'' {\em Quantum
  Science and Technology}, vol.~3, no.~1, p.~013001, 2018.

\bibitem{OBrien2010}
J.~L. O'Brien, A.~Furusawa, and J.~Vu{\v{c}}kovi{\'{c}}, ``{Photonic quantum
  technologies},'' {\em Nat. photon.}, vol.~3, no.~12, p.~687, 2009.

\bibitem{Aharonovich2016}
I.~Aharonovich, D.~Englund, and M.~Toth, ``{Solid-state single-photon
  emitters},'' {\em Nat. photon.}, vol.~10, no.~10, p.~631, 2016.

\bibitem{He2013}
Y.-M. He, Y.~He, Y.-J. Wei, D.~Wu, M.~Atat{\"{u}}re, C.~Schneider,
  S.~H{\"{o}}fling, M.~Kamp, C.-Y. Lu, and J.-W. Pan, ``{On-demand
  semiconductor single-photon source with near-unity indistinguishability},''
  {\em Nat. Nanotechnol.}, vol.~8, no.~February, p.~213, 2013.

\bibitem{Arcari2014}
M.~Arcari, I.~S{\"{o}}llner, A.~Javadi, S.~{Lindskov Hansen}, S.~Mahmoodian,
  J.~Liu, H.~Thyrrestrup, E.~H. Lee, J.~D. Song, S.~Stobbe, and P.~Lodahl,
  ``{Near-Unity Coupling Efficiency of a Quantum Emitter to a Photonic Crystal
  Waveguide},'' {\em Phys. Rev. Lett.}, vol.~113, no.~9, p.~093603, 2014.

\bibitem{Sapienza2015}
L.~Sapienza, M.~Davan{\c{c}}o, A.~Badolato, and K.~Srinivasan, ``{Nanoscale
  optical positioning of single quantum dots for bright and pure single-photon
  emission},'' {\em Nat. commun.}, vol.~6, p.~7833, 2015.

\bibitem{Somaschi2016}
N.~Somaschi, V.~Giesz, L.~{De Santis}, J.~C. Loredo, M.~P. Almeida,
  G.~Hornecker, S.~L. Portalupi, T.~Grange, C.~Ant{\'{o}}n, J.~Demory,
  C.~Gom{\'{e}}z, I.~Sagnes, N.~D. Lanzillotti-Kimura, A.~Lema{\'{i}}tre,
  A.~Auffeves, A.~G. White, L.~Lanco, and P.~Senellart, ``{Near-optimal
  single-photon sources in the solid state},'' {\em Nat. photon.}, vol.~10,
  no.~2, p.~340, 2016.

\bibitem{Ding2016}
X.~Ding, Y.~He, Z.~C. Duan, N.~Gregersen, M.~C. Chen, S.~Unsleber, S.~Maier,
  C.~Schneider, M.~Kamp, S.~H{\"{o}}fling, C.-Y. Lu, and J.-W. Pan,
  ``{On-Demand Single Photons with High Extraction Efficiency and Near-Unity
  Indistinguishability from a Resonantly Driven Quantum Dot in a
  Micropillar},'' {\em Phys. Rev. Lett.}, vol.~116, no.~2, p.~020401, 2016.

\bibitem{Daveau2016}
R.~S. Daveau, K.~C. Balram, T.~Pregnolato, J.~Liu, E.~H. Lee, J.~D. Song,
  V.~Verma, R.~Mirin, S.~W. Nam, L.~Midolo, S.~Stobbe, K.~Srinivasan, and
  P.~Lodahl, ``{Efficient fiber-coupled single-photon source based on quantum
  dots in a photonic-crystal waveguide},'' {\em Optica}, vol.~4, no.~2, p.~178,
  2016.

\bibitem{Kirsanske2017}
G.~Kir\ifmmode \check{s}\else \v{s}\fi{}ansk\ifmmode~\dot{e}\else \.{e}\fi{},
  H.~Thyrrestrup, R.~S. Daveau, C.~L. Dree\ss{}en, T.~Pregnolato, L.~Midolo,
  P.~Tighineanu, A.~Javadi, S.~Stobbe, R.~Schott, A.~Ludwig, A.~D. Wieck, S.~I.
  Park, J.~D. Song, A.~V. Kuhlmann, I.~S\"ollner, M.~C. L\"obl, R.~J.
  Warburton, and P.~Lodahl, ``Indistinguishable and efficient single photons
  from a quantum dot in a planar nanobeam waveguide,'' {\em Phys. Rev. B},
  vol.~96, p.~165306, Oct 2017.

\bibitem{Kuhlmann2013}
A.~V. Kuhlmann, J.~Houel, A.~Ludwig, L.~Greuter, D.~Reuter, A.~D. Wieck,
  M.~Poggio, and R.~J. Warburton, ``{Charge noise and spin noise in a
  semiconductor quantum device},'' {\em Nat. physics}, vol.~9, no.~9, p.~570,
  2013.

\bibitem{Urbaszek2013}
B.~Urbaszek, X.~Marie, T.~Amand, O.~Krebs, P.~Voisin, P.~Maletinsky,
  A.~H{\"{o}}gele, and A.~Imamoglu, ``{Nuclear spin physics in quantum dots: An
  optical investigation},'' {\em Rev. Mod. Phys.}, vol.~85, p.~79, jan 2013.

\bibitem{Besombes2001}
L.~Besombes, K.~Kheng, L.~Marsal, and H.~Mariette, ``{Acoustic phonon
  broadening mechanism in single quantum dot emission},'' {\em Phys. Rev. B},
  vol.~63, p.~155307, 2001.

\bibitem{Kuhlmann2015}
A.~V. Kuhlmann, J.~H. Prechtel, J.~Houel, A.~Ludwig, D.~Reuter, A.~D. Wieck,
  and R.~J. Warburton, ``{Transform-limited single photons from a single
  quantum dot},'' {\em Nat. commun.}, vol.~6, p.~8204, 2015.

\bibitem{Lobl2017}
M.~C. L{\"{o}}bl, I.~S{\"{o}}llner, A.~Javadi, T.~Pregnolato, R.~Schott,
  L.~Midolo, A.~V. Kuhlmann, S.~Stobbe, A.~D. Wieck, P.~Lodahl, A.~Ludwig, and
  R.~J. Warburton, ``{Narrow optical linewidths and spin pumping on
  charge-tunable close-to-surface self-assembled quantum dots in an ultrathin
  diode},'' {\em Phys. Rev. B}, vol.~96, p.~165440, 2017.

\bibitem{Thyrrestrup2018}
H.~Thyrrestrup, G.~Kir{\v{s}}ansk\ifmmode~\dot{e}\else \.{e}\fi{}, H.~{Le
  Jeannic}, T.~Pregnolato, L.~Zhai, L.~Raahauge, L.~Midolo, N.~Rotenberg,
  A.~Javadi, R.~Schott, A.~D. Wieck, A.~Ludwig, M.~C. L{\"{o}}bl,
  I.~S{\"{o}}llner, R.~J. Warburton, and P.~Lodahl, ``{Quantum Optics with
  Near-Lifetime-Limited Quantum-Dot Transitions in a Nanophotonic Waveguide},''
  {\em Nano Lett.}, vol.~18, no.~3, p.~1801, 2018.

\bibitem{Borri2001}
P.~Borri, W.~Langbein, S.~Schneider, U.~Woggon, R.~L. Sellin, D.~Ouyang, and
  D.~Bimberg, ``{Ultralong Dephasing Time in InGaAs Quantum Dots},'' {\em Phys.
  Rev. Lett.}, vol.~87, no.~15, p.~157401, 2001.

\bibitem{Bayer2002}
M.~Bayer and A.~Forchel, ``{Temperature dependence of the exciton homogeneous
  linewidth in In$_{0.60}$Ga$_{0.40}$As/GaAs self assembled quantum dots},''
  {\em Phys. Rev. B}, vol.~65, no.~4, p.~041308, 2002.

\bibitem{Krummheuer2002}
B.~Krummheuer, V.~M. Axt, and T.~Kuhn, ``{Theory of pure dephasing and the
  resulting absorption line shape in semiconductor quantum dots},'' {\em Phys.
  Rev. B}, vol.~65, no.~19, p.~195313, 2002.

\bibitem{Mahan}
G.~D. Mahan, {\em {Many-Particle Physics}}.
\newblock {Springer US}, 2000.

\bibitem{Muljarov2004}
E.~A. Muljarov and R.~Zimmermann, ``{Dephasing in Quantum Dots: Quadratic
  Coupling to Acoustic Phonons},'' {\em Phys. Rev. Lett.}, vol.~93, no.~23,
  p.~237401, 2004.

\bibitem{Lindwall2007}
G.~Lindwall, A.~Wacker, C.~Weber, and A.~Knorr, ``{Zero-Phonon Linewidth and
  Phonon Satellites in the Optical Absorption of Nanowire-Based Quantum
  Dots},'' {\em Phys. Rev. Lett.}, vol.~99, p.~087401, 2007.

\bibitem{Tighineanu2018}
P.~Tighineanu, C.~L. Dree{\ss}en, C.~Flindt, P.~Lodahl, and A.~S. S{\o}rensen,
  ``{Phonon Decoherence of Quantum Dots in Photonic Structures: Broadening of
  the Zero-Phonon Line and the Role of Dimensionality},'' {\em
  arXiv:1702.04812v2}, 2018.

\bibitem{Iles-Smith2017a}
J.~Iles-Smith, D.~P.~S. McCutcheon, A.~Nazir, and J.~M{\o}rk, ``{Phonon
  scattering inhibits near-unity efficiency and indistinguishability in
  semiconductor single-photon sources},'' {\em Nat. photon.}, vol.~11, no.~8,
  p.~521, 2017.

\bibitem{Massimo1996}
G.~M. Palma, K.-A. Suominen, and A.~K. Ekert, ``Quantum computers and
  dissipation,'' {\em Proceedings: Mathematical, Physical and Engineering
  Sciences}, vol.~452, no.~1946, pp.~567--584, 1996.

\bibitem{Galland2008}
C.~Galland, A.~H\"ogele, H.~E. T\"ureci, and A.~Imamo\ifmmode~\breve{g}\else
  \u{g}\fi{}lu, ``Non-markovian decoherence of localized nanotube excitons by
  acoustic phonons,'' {\em Phys. Rev. Lett.}, vol.~101, p.~067402, Aug 2008.

\bibitem{Takaghara1999}
T.~Takagahara, ``{Theory of exciton dephasing in semiconductor quantum dots},''
  {\em Phys. Rev. B}, vol.~60, no.~4, p.~2638, 1999.

\bibitem{parameters}
Parameters for the GaAs waveguide: $\rho =
  \SI{5317}{\kilogram\per\meter^3}$\cite{Straumanis1965}, $E =
  \SI{87}{GPa}$\cite{Cottam2001}, $\nu = 0.31$\cite{Cottam2001}, $n =
  3.5$\cite{Lodahl2015}, $v_\text{s} =
  \SI{4770}{\meter\per\second}$\cite{Tighineanu2018}, $L_\text{QD} =
  \SI{3}{\nano\meter}$\cite{Muljarov2004}, $D_e =
  -\SI{14.6}{\electronvolt}$\cite{Lindwall2007}, $D_h =
  -\SI{4.8}{\electronvolt}$\cite{Lindwall2007}, $\Delta_{e} = 2\Delta_{h} =
  -\SI{40}{\milli\electronvolt}$\cite{Tighineanu2018}, $W =
  \SI{300}{\nano\meter}$\cite{Kirsanske2017}, $H =
  \SI{175}{\nano\meter}$\cite{Kirsanske2017}.\\ Paramters for SiO$_2$: $\rho =
  \SI{2200}{\kilogram\per\meter^3}$\cite{Chiou2007}, $E =
  \SI{75}{GPa}$\cite{Chiou2007}, $\nu = 0.17$\cite{Chiou2007}, $n =
  1.45$\cite{Malitson1965}.

\bibitem{Mason1966}
W.~P. Mason, ed., {\em Physical Acoustics: Principles and Methods}, vol.~4.
\newblock {Academic Press}, 1966.

\bibitem{Luxmoore2013}
I.~J. Luxmoore, N.~A. Wasley, A.~J. Ramsay, A.~C.~T. Thijssen, R.~Oulton,
  M.~Hugues, S.~Kasture, V.~G. Achanta, A.~M. Fox, and M.~S. Skolnick,
  ``{Interfacing Spins in an InGaAs Quantum Dot to a Semiconductor Waveguide
  Circuit Using Emitted Photons},'' {\em Phys. Rev. Lett.}, vol.~110, no.~3,
  p.~037402, 2013.

\bibitem{Faraon2008}
A.~Faraon, I.~Fushman, D.~Englund, N.~Stoltz, P.~Petroff, and
  J.~Vu{\v{c}}kovi{\'{c}}, ``{Dipole induced transparency in waveguide coupled
  photonic crystal cavities},'' {\em Optics Express}, vol.~16, p.~12154, 2008.

\bibitem{Javadi2015}
A.~Javadi, I.~S{\"{o}}llner, M.~Arcari, S.~{Lindskov Hansen}, L.~Midolo,
  S.~Mahmoodian, G.~Kir{\v{s}}ansk\ifmmode~\dot{e}\else \.{e}\fi{},
  T.~Pregnolato, E.~H. Lee, J.~D. Song, S.~Stobbe, and P.~Lodahl,
  ``{Single-photon non-linear optics with a quantum dot in a waveguide},'' {\em
  Nat. commun.}, vol.~6, p.~8655, 2015.

\bibitem{Laucht2012}
A.~Laucht, S.~P{\"{u}}tz, T.~G{\"{u}}nthner, N.~Hauke, R.~Saive,
  S.~Fr{\'{e}}d{\'{e}}rick, M.~Bichler, M.~C. Amann, A.~W. Holleitner,
  M.~Kaniber, and J.~J. Finley, ``{A waveguide-Coupled On-Chip Single-Photon
  Source},'' {\em Phys. Rev. X}, vol.~2, no.~1, p.~011014, 2012.

\bibitem{Fan1999}
R.~S. Fan and R.~{Brian Hooker}, ``{Tapered Polymer Single-Mode Waveguides for
  Mode Transformation},'' {\em Journal of Lightwave Technology}, vol.~17,
  no.~3, pp.~466--474, 1999.

\bibitem{McNab2003}
S.~McNab, N.~Moll, and Y.~Vlasov, ``{Ultra-low loss photonic integrated circuit
  with membrane-type photonic crystal waveguides},'' {\em Optics Express},
  vol.~11, no.~22, p.~2927, 2003.

\bibitem{Roelkens2006}
G.~Roelkens, D.~{Van Thourhout}, R.~Baets, R.~N{\"{o}}tzel, and M.~Smit,
  ``{Laser emission and photodetection in an InP/InGaAsP layer integrated on
  and coupled to a Silicon-on-Insulator waveguide circuit},'' {\em Optics
  Express}, vol.~14, no.~18, p.~8154, 2006.

\bibitem{Midolo2015a}
L.~Midolo, T.~Pregnolato, G.~Kir{\v{s}}ansk\ifmmode~\dot{e}\else \.{e}\fi{},
  and S.~Stobbe, ``Soft-mask fabrication of gallium arsenide nanomembranes for
  integrated quantum photonics,'' {\em Nanotechnology}, vol.~26, no.~48,
  p.~484002, 2015.

\bibitem{Straumanis1965}
M.~E. Straumanis and C.~D. Kim, ``{Phase extent of gallium arsenide determined
  by the lattice constant and density method},'' {\em Acta Crystallographica},
  vol.~19, no.~2, pp.~256--259, 1965.

\bibitem{Cottam2001}
R.~I. Cottam and G.~A. Saunders, ``{The elastic constants of GaAs from 2 K to
  320 K},'' {\em Journal of Physics C: Solid State Physics}, vol.~6, no.~13,
  pp.~2105--2118, 2001.

\bibitem{Chiou2007}
D.~Y. Chiou, M.~Y. Chen, M.~W. Chang, and H.~C. Deng, ``{Characterization and
  optimization design of the polymer-based capacitive micro-arrayed ultrasonic
  transducer},'' {\em Japanese Journal of Applied Physics, Part 1: Regular
  Papers and Short Notes and Review Papers}, vol.~46, no.~11, pp.~7496--7503,
  2007.

\bibitem{Malitson1965}
I.~H. Malitson, ``{Interspecimen Comparison of the Refractive Index of Fused
  Silica},'' {\em Journal of the Optical Society of America}, vol.~55, no.~10,
  p.~1205, 1965.

\end{thebibliography}

\end{document}